\documentclass[12pt]{iopart}
\usepackage{graphicx}
\usepackage{wrapfig}

\begin{document}

\title{$\Upsilon$ production in p+p and Au+Au collisions in STAR}

\author{Debasish Das for the STAR Collaboration}

\address{Physics Department, University of California, Davis, CA, 95616 USA.}
\ead{debasish@rcf.rhic.bnl.gov, debdas@ucdavis.edu}
\begin{abstract}

The study of quarkonium production in relativistic heavy ion collisions provides insight into the properties 
of the produced medium. The lattice studies show a sequential suppression of quarkonia states when compared to normal nuclear 
matter; which further affirms that a full spectroscopy including bottomonium can provide us a better thermometer for the matter produced 
under extreme conditions in relativistic heavy ion collisions. With the completion of the STAR Electromagnetic Calorimeter and with the 
increased luminosity provided by RHIC in Run 6 and 7, the study of $\Upsilon$ production via the di-electron channel becomes possible. 
We present the results on $\Upsilon$ measurements in p+p collisions (from Run 6) along with the first results 
from Au+Au collisions (in Run 7)  at $\sqrt{s_{\rm{NN}}} = 200$ GeV from the STAR experiment.

\end{abstract}


\section{Introduction}

One of the definitive predictions of Quantum Chromodynamics (QCD) is that at sufficiently high density and temperature, 
strongly interacting matter will be in a deconfined state of the quarks and gluons, called the Quark-Gluon Plasma (QGP).
In a QGP the strong interactions become screened. The production of heavy quarkonia states in p+p, p+A and A+A collisions 
provide an important tool to study the properties of QGP~\cite{Satz:1995sa}. The larger production cross-sections for charmonium~\cite{Adler:2003rc}
states compared to bottomonium states have initiated the studies of charmonium along with
the observation of charmonium suppression~\cite{Matsui:1986dk,Arnaldi:2007aa} in relativistic heavy ion collisions.
The high energy at RHIC allows the measurement of the $\Upsilon$ states in heavy ion collisions, where the 
lattice QCD studies of quarkonia spectral functions suggest that while
the $\Upsilon''$ melts at RHIC and the $\Upsilon'$ is likely to melt,
the $\Upsilon$ is expected to survive~\cite{Digal:2001ue,Wong:2004zr}.
Their suppression pattern can be used as a thermometer to the QCD matter~\cite{Grandchamp:2005yw} and henceforth the 
study of $\Upsilon$ production is of paramount importance.

As the bottomonium state is massive ($\sim$ 10 GeV/$c^{2}$) its decay leptons have sufficiently large
momenta and bottomonium spectroscopy requires higher luminosities. Its decay leptons have sufficiently
large momenta above the background processes which helps in high-level triggering even in central Au+Au collisions. 
In this paper, we report preliminary $\Upsilon$ measurements in p+p collisions (from Run 6) along with preliminary results 
from Au+Au collisions (in Run 7) at mid-rapidity obtained with the STAR detector.

\section{Experimental Setup}

The golden quarkonium decay mode for STAR is $\Upsilon\rightarrow e^+e^-$. The 
main detectors for this analysis are the TPC (Time Projection Chamber)~\cite{Anderson:2003ur} and 
the BEMC (Barrel Electro-Magnetic Calorimeter)~\cite{Beddo:2002zx}. The advantages of STAR are its 
large acceptance along with the trigger capabilities of the BEMC and combined electron 
identification using the TPC+BEMC. The suitable triggers using the BEMC allows us to suppress 
the hadrons, achieving a higher signal to background ratio. Thus we can trigger 
on electron-positron pairs with a given invariant mass at sufficiently high rates.

\section{The STAR Quarkonia Trigger Setup}

The STAR $\Upsilon$ trigger is a two-stage setup which comprises of a fast  
level-0 (L0) hardware component ($\sim 1~\mu$s) and a level-2 (L2) software component
($\sim 100~\mu$s).
The L0 trigger is a fast hardware trigger taking a decision for each RHIC bunch crossing and 
consisting of a four layer tree structure of data storage and manipulation boards (DSM). 
Such a trigger is issued if at least one BEMC tower is above the adjusted quarkonium threshold of 
$E_{T} > $ 3.5 GeV and the associated trigger patch having a total energy above 4.3 GeV, along with a minimum bias condition.
The L2 trigger is a software trigger which analyses events at the rate of about 1 kHz. In the initial 
step L2 starts finding towers with similar L0 threshold. It performs clustering to account for energy leaking into the adjacent towers, 
and to improve the position resolution by calculating the mean cluster position, weighted by the energy 
seen in each tower. Cuts are applied on the invariant mass
$m_{ee}=\sqrt{2E_1E_2(1-\cos\theta_{12})}$ and the opening angle $\theta_{12}$ between clusters. 
Thus it aborts the read-out of all detectors if the algorithm does not detect at least one pair with the 
invariant mass within a given mass window. The trigger though limited by dead-time of the data acquisition system (DAQ) 
was utilized efficiently for p+p (in 2006) and Au+Au collisions (in 2007). Thus enabling to process the complete data-set covering 
the total luminosity provided by RHIC, in just a few weeks after the completion of physics Run.

\section{$\Upsilon$ Analysis and Results for p+p and Au+Au collisions}

\begin{figure}
\begin{minipage}[b]{.46\linewidth}
\includegraphics*[width=7cm]{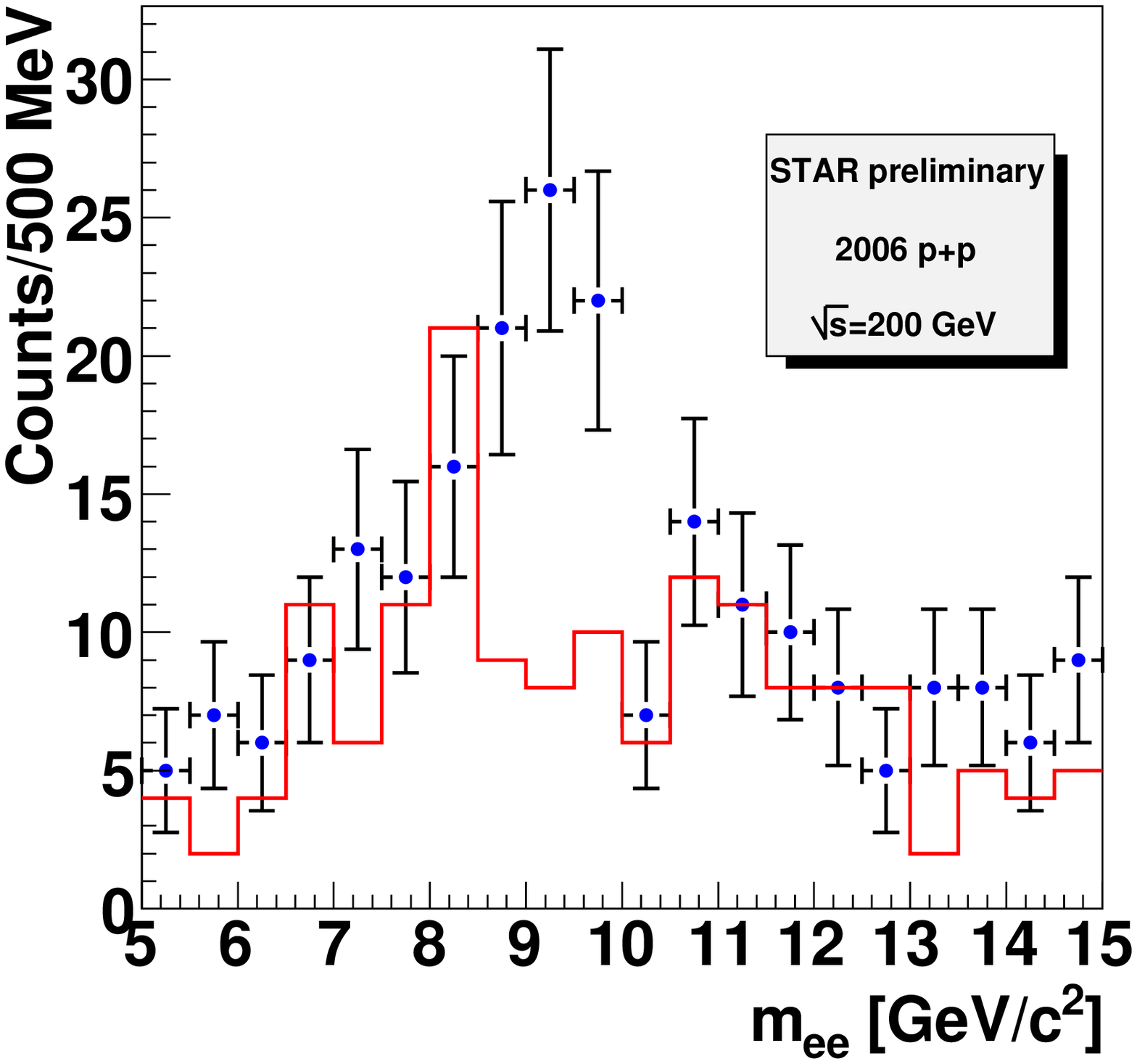}
\end{minipage}\hfill
\vspace{-2mm}
\begin{minipage}[b]{.46\linewidth}
\includegraphics*[width=7cm]{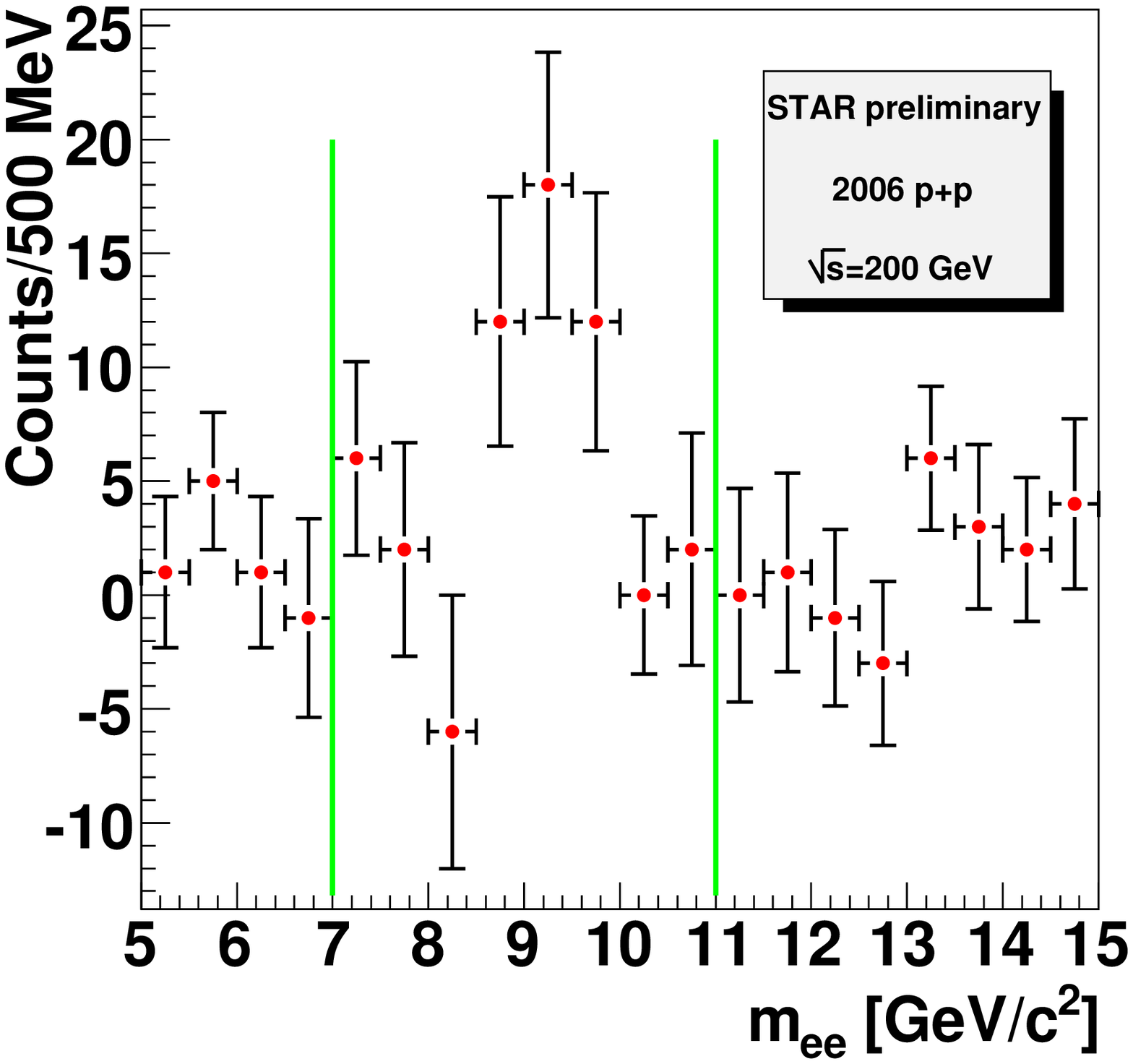}
\end{minipage}
\vspace{-2mm}
\caption{Left panel: STAR 2006 p+p collisions $\Upsilon \rightarrow e^+e^-$
at $\sqrt{s} = 200$ GeV signal and background  with statistical error bars
from combining unlike-sign pairs (electrons and positrons). The red background results
from combining like-sign pairs. Right panel: Background-subtracted $\Upsilon$ signal with
statistical error bars. The green vertical bars mark the boundaries of integration for the yield.}
\label{fig:upsilon_peak06}
\end{figure}

For p+p collisions at $\sqrt{s} = 200$ GeV in 2006, with the full BEMC acceptance, STAR sampled $\sim$ 9 pb$^{-1}$ of integrated luminosity. 
Two different trigger setups were deployed. The preliminary analysis focused on one trigger setup with
the integrated luminosity $\int\mathcal{L}dt \sim$ 5.6 pb$^{-1}$ was reported in Ref~\cite{Djawotho:2007mj}.
Electrons were identified by selecting charged particle tracks where number of fitted points are 
greater than 20 out of 45, along with the specific $dE/dx$ ionization energy loss in the TPC that deposited more than 3 GeV of energy in a BEMC tower.
Electron-positron pairs were then combined to produce the invariant mass spectrum.
Finally the like-sign electron pairs were combined to form the invariant mass spectrum of
the background which was subtracted from the unlike-sign spectrum as shown in Fig.~\ref{fig:upsilon_peak06}~\cite{Djawotho:2007mj}.

\begin{figure}
\begin{minipage}[b]{.46\linewidth}
\includegraphics*[width=7cm]{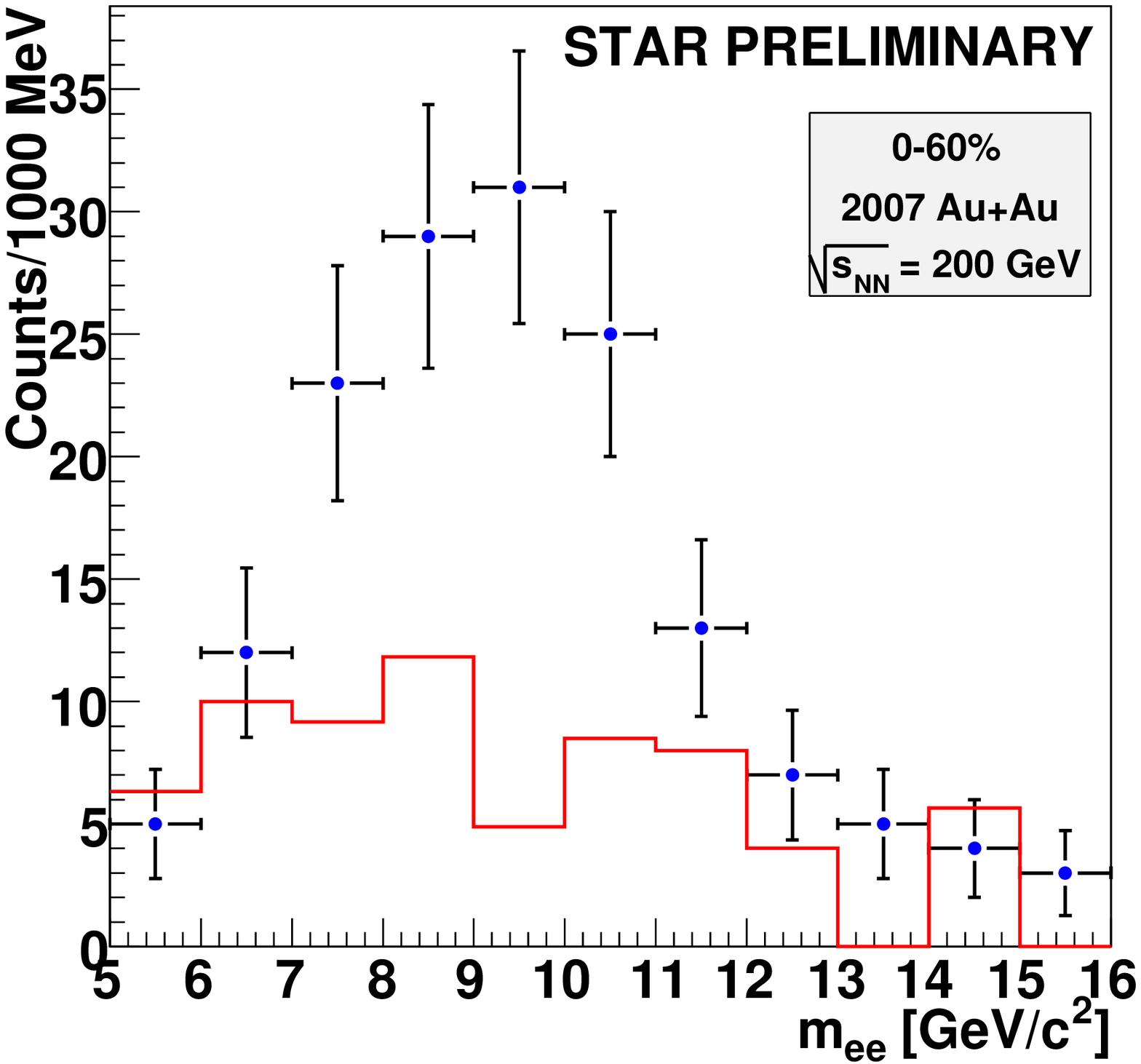}
\end{minipage}\hfill
\vspace{-2mm}
\begin{minipage}[b]{.46\linewidth}
\includegraphics*[width=7cm]{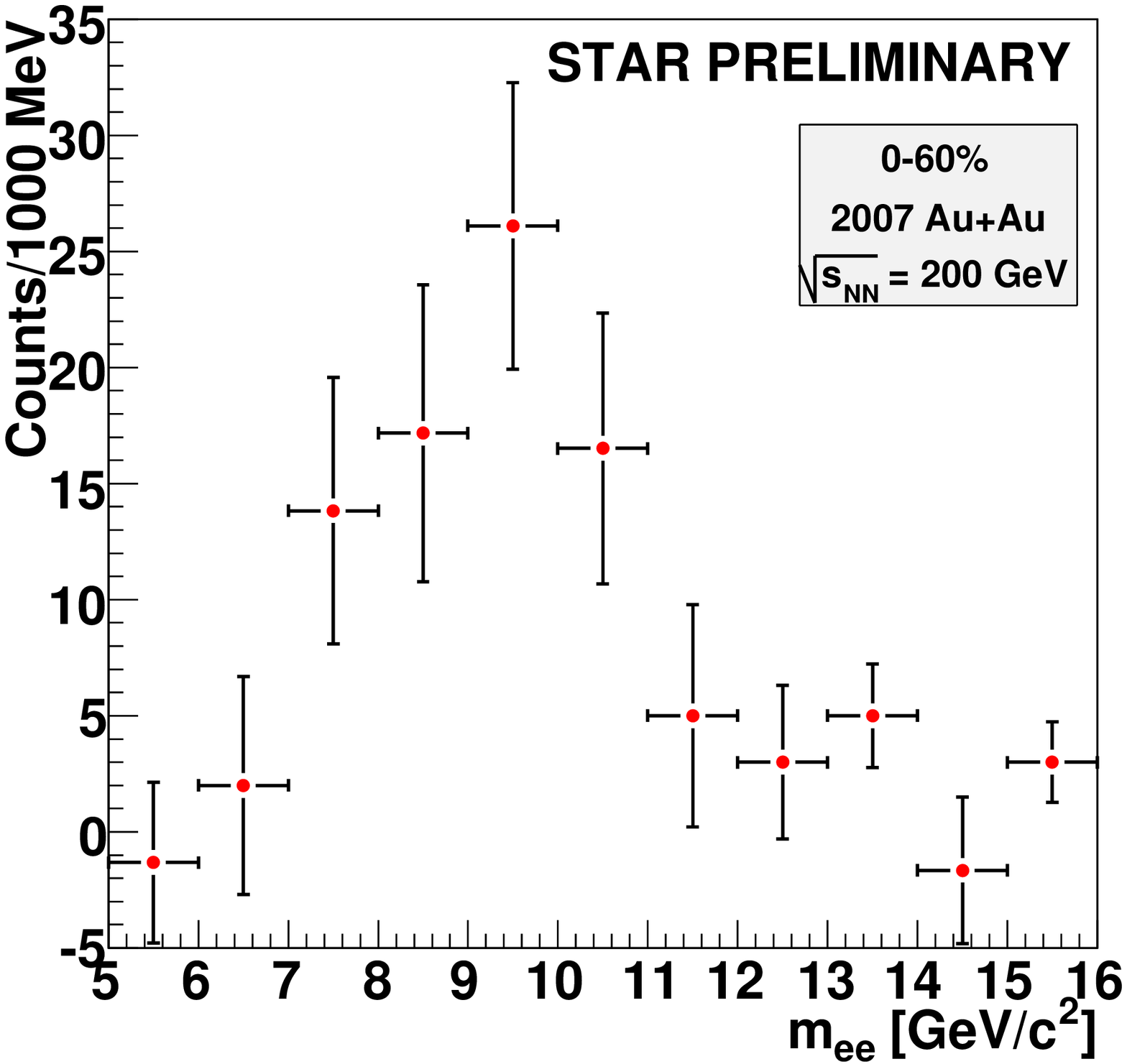}
\end{minipage}
\vspace{-2mm}
\caption{Left panel: STAR 2007 Au+Au for 0-60$\%$ central collisions $\Upsilon \rightarrow e^+e^-$
at $\sqrt{s_{\rm{NN}}} = 200$ GeV signal and background  with statistical error bars
from combining unlike-sign pairs (electrons and positrons). The red background results
from combining like-sign pairs. Right panel: Background-subtracted $\Upsilon$ signal with
statistical error bars.}
\label{fig:upsilon_peak07}
\end{figure}

Since its not possible to resolve the individual states of the $\Upsilon$ family with the 
available statistics, the yield reported here is for the combined
$\Upsilon+\Upsilon'+\Upsilon''$ states. The total yield was extracted by integrating
the invariant mass spectrum from 7 to 11 GeV/$c^2$ as shown by the vertical boundaries 
in the right panel of Fig.~\ref{fig:upsilon_peak06}. 
The width of the peak $\sim$ 1 GeV/$c^2$ was found to be consistent with simulation. 
The significance of the signal was estimated at $3\sigma$. The estimated contribution from Drell-Yan was $\sim 9\%$ based on PYTHIA. 
We find for the cross section at mid-rapidity in $\sqrt{s} = 200$ GeV p+p collisions
$BR\times(d\sigma/dy)^{\Upsilon+\Upsilon'+\Upsilon''}_{y=0}=91\pm 28~{\rm (stat.)}
\pm 22~{\rm (syst.)~pb}$~\cite{Djawotho:2007mj}. The systematic error is dominated by the uncertainty in the integrated luminosity.

In 2007, Au+Au collisions at $\sqrt{s_{\rm{NN}}} = 200$ GeV, with the total BEMC acceptance, STAR sampled 300 $\mu$b$^{-1}$ of integrated luminosity. 
Two different trigger setups were used but the preliminary analysis in Au+Au collisions was centered on one trigger setup with larger 
integrated luminosity of $\int\mathcal{L}dt \sim 262 ~\mu$b$^{-1}$. The electron identification conditions were 
kept close to what we had used for p+p 2006 analysis. The like-sign electron pairs were combined to form the invariant mass spectrum of
the background which was subtracted from the unlike-sign spectrum presented in Fig.~\ref{fig:upsilon_peak07}.
The significance of the signal was estimated at $4\sigma$ for 0-60$\%$ centrality in Au+Au collisions at $\sqrt{s_{\rm{NN}}} = 200$ GeV.
The upsilon line shape in Au+Au could be wider than in p+p and this is under study. 
The trigger efficiency and systematic checks are in progress towards the estimation of the nuclear
modification factor, $R_{AA}$, integrated over the transverse momenta and 0-60$\%$ centrality.

\section{Conclusions}

The STAR experiment made the first RHIC measurement of the
$\Upsilon,\Upsilon',\Upsilon''\rightarrow e^+e^-$ cross section at mid-rapidity in p+p collisions
at $\sqrt{s} = 200$ GeV; $BR\times(d\sigma/dy)_{y=0}=91\pm28~{\rm (stat.)}\pm22~{\rm (syst.)}$.
The STAR $\Upsilon$ measurement is consistent with the world data and NLO in the CEM
(Color Evaporation Model) pQCD calculations~\cite{Djawotho:2007mj,rhic2}.
The full BEMC acceptance and suitable trigger setups are essential for a successful quarkonia program at STAR.
The first ever preliminary measurements for $\Upsilon$ invariant mass in Au+Au collisions at $\sqrt{s_{\rm{NN}}} = 200$ GeV 
are discussed and presented. The comparative study of p+p and Au+Au data-sets towards the nuclear modification factor is in progress. 
To get a further understanding of the cold nuclear matter effects, the d+Au collisions at $\sqrt{s_{\rm{NN}}} = 200$ GeV was taken in Run 8 and 
will be analysed soon.

\end{document}